\documentclass[letterpaper]{jpconf}

\usepackage{graphicx}

\begin{document}

\title{From Microscales to Macroscales in 3D: Selfconsistent Equation of State for Supernova and Neutron Star Models}

\author{W G Newton$^1$, J R Stone$^{1,2,3}$, A Mezzacappa$^2$}

\address{$^1$ Department of Physics, University of Oxford, Oxford OX1 3PU, United Kingdom}
\address{$^2$ Physics Division, Oak Ridge National Laboratory,P.O. Box 2008, Oak Ridge, TN 37831, USA}
\address{$^3$ Department of Chemistry and Biochemistry, University of Maryland, College Park, MD 20742, USA}

\ead{william.newton@seh.ox.ac.uk, stonejr@ornl.gov, mezzacappaa@ornl.gov}

\begin{abstract}
First results from a fully self-consistent, temperature-dependent
equation of state that spans the whole density range of neutron
stars and supernova cores are presented. The equation of state
(EoS) is calculated using a mean-field Hartree-Fock method in
three dimensions (3D). The nuclear interaction is represented by
the phenomenological Skyrme model in this work, but the EoS can be
obtained in our framework for any suitable form of the
nucleon-nucleon effective interaction. The scheme we employ
naturally allows effects such as (i) neutron drip, which results
in an external neutron gas, (ii) the variety of exotic nuclear
shapes expected for extremely neutron heavy nuclei, and (iii) the
subsequent dissolution of these nuclei into nuclear matter. In
this way, the equation of state is calculated across phase
transitions without recourse to interpolation techniques between
density regimes described by different physical models. EoS tables
are calculated in the wide range of densities, temperature and
proton/neutron ratios on the ORNL NCCS XT3, using up to 2000
processors simultaneously.
\end{abstract}

\section{Introduction}

Observational properties of neutron stars and supernovae serve as
powerful constraints of the nuclear EoS. There is a large variety
of EoS models in the literature and it is imperative to
investigate the connection of the physical processes expected in
stars with the features of individual EoS. The models used to
construct nuclear EoS range from empirical to those based on
non-relativistic effective and realistic nucleon-nucleon
potentials and relativistic field theories (for a recent reviews
see e.g. \cite{lat00,sto06}). It is unclear at present which of
these EoS is closest to reality. All the EoS are required to
reflect physics occurring in a wide region of particle number
densities. In core-collapse supernovae these densities span from
the subnuclear density of about 4$\times$10$^{-8}$ to $\sim$0.1
fm$^{-3}$ (inhomogeneous matter) to the high density phase
(uniform matter) between $\sim$0.1 fm$^{-3}$ -- 0.6 fm$^{-3}$.
Neutron star models involve an even wider density range starting
from $\sim$6$\times$10$^{-15}$ fm$^{-3}$ (estimated density at the
surface of neutron stars) to about 0.6--1.0 fm$^{-3}$ (expected in
the center of neutron stars). Most of the currently used EoSs in
both the subnuclear and supernuclear density do not cover the
whole range but are composed of several EoSs reflecting the
evolution of the character of matter with changing density in
smaller intervals. One of the most interesting density regions
covers the transition between uniform and inhomogenous matter,
known is the `pasta' phase. In this region superheavy neutron rich
nuclei beyond the neutron drip line gradually dissolve into
nucleon + lepton matter of uniform density. The proton and neutron
density distribution is determined by a delicate balance between
the surface tension of nuclei and the Coulomb repulsion of
protons. Previous models of the `pasta' phase of matter, assuming
spherical symmetry, predicted the existence of a series of exotic
nuclear shapes - rods, slabs, tubes and bubbles, immersed in a
free neutron and electron gas, corresponding to minimal energy of
the matter as a function of increasing density, until the uniform
distribution becomes energetically favorable. The 'pasta' phase of
stellar matter, although occurring in a relatively small region of
density, has a significant influence on the neighboring higher and
lower density regions due to the requirement of continuity and
thermodynamical consistency of the energy per particle and related
quantities throughout the whole density and temperature range.

The focus of this work is on the EoS that serves as an input to
core-collapse supernova models and non-equilibrium young neutron
stars. However, only a slight modification, i.e. the inclusion of
chemical equilibrium at supernuclear densities, is required to use
this EoS in old neutron stars. The most widely used EoS in
core-collapse supernova simulations so far have been the
non-relativistic EoS by Lattimer-Swesty \cite{lat91} and
relativistic mean-field model by Shen et al \cite{she98}. Both
these EoS describe hot stellar matter assuming spherical symmetry
and use different models for matter in different density and
temperature regions. It is the aim of this work to show that a
fully self-consistent non-relativistic EoS in the Hartree-Fock
(HF) approximation \cite{hartree,fock} in three dimensions
(removing the constraint of spherical symmetry) can be constructed
in the whole density and temperature region of interest. In this
way the matter is treated as an ensemble of nucleons that
naturally configure to a distribution corresponding to the minimal
energy per particle at given density and temperature. The
computation method adopted here is an extension of previous work
of Bonche and Vautherin \cite{bon81} and Hillebrant and Wolff
\cite{hil85} who calculated self-consistent HF EoS at finite
temperature but only in the spherically symmetrical case and
Magierski and Heenen \cite{mag02} who developed an HF EoS for the
general case of three dimensions but considered only zero
temperature.

\section{Computational Procedure}
Equation of State, determining the pressure of a system as a
function of density and temperature, is constructed for stellar
matter at the densities and temperatures found during core
collapse of a massive star pre- and post-bounce. Such matter is
composed primarily of neutrons, protons and electrons, with a
significant flux of photons, positrons and neutrinos also present
during core collapse. There are three main bulk parameters of the
matter, baryon number density $n_{\rm b}$, temperature T and
proton fraction $y_{\rm p}$ defined as the ratio of the proton
number density $n_{\rm p}$ to the total baryon number density
$n_{\rm b}$. In the present work, the ranges of these parameters
are $0.001 < n_b < 0.16 fm^{-3}$, $0 < T < 10 MeV$, and $0 <
y_{\rm p} < 0.5$. Furthermore, the EoS is dependent on a number of
microscopic parameters, determining the strong force, acting
between nucleons in the matter. The phenomenological Skyrme
SkM$^*$ force \cite{skyrme} is used here but it is easy to modify
the computer code for any other applicable model of the
nucleon-nucleon interaction. Finally, the electric Coulomb force
acting between charged particles, protons and electrons is
included. Electrons are treated as forming a degenerate Fermi gas
which should be a valid approximation. Neutrinos are not
considered at the present stage of the model.

The fundamental assumption used here is that nuclear matter has a
periodic character and can be modeled as an infinite sequence of
cubic unit cells. This notion removes a serious limitation of all
previous models based on consideration of spherical cells which
allows only spherically symmetrical nucleon distribution in the
cell and cannot fully express the period character of matter as
the cells make contact only at limited number of points leaving
the space between them unaccounted for. Each unit cell contains a
certain number of neutrons $N$ and protons $Z$, making a total
baryon number of $A = N + Z$. Quantum mechanical determination of
all energy states and corresponding wave functions of a system of
A nucleons in the cell requires exact solution of the
A-dimensional equation of motion - the Schroedinger Equation -
which is not technically feasible at present. However, if it is
assumed that there exists an average single-particle potential,
created by all nucleons, in which each nucleon moves independently
of all the other nucleons present, then it is possible to use the
Hartree-Fock approximation to the A-dimensional problem which
reduces it to A one-dimensional problems.A spectrum of discrete
energy states, the single-particle states, can be defined in the
cell which the individual nucleons occupy (in analogy to a
spectrum of standing waves in a box in classical physics). The
single-particle wave functions $\psi_i$, associated with these
states, are used to construct the total wavefunction  $\Psi$ and
to calculate the expectation value of total energy in the state
$\Psi$. Obviously there are many ways the nucleons can be
distributed over the available single-particle states, which
always considerably outnumber, by a factor of two at least, the
the total number of nucleons in the cell. Each of these nucleon
configurations corresponds to an energy state and a particular
spacial distribution of nucleon density in the cell. It turns out
that it is possible to find a state $\Psi_{\rm_{min}}$,
constructed of a set of single-particle states, of which the
lowest A states are occupied, which corresponds to the minimum
energy of the system and is the best approximation to the true
A-particle ground state.

Starting from a trial set of single-particle wave functions
$\psi_i$, the expectation value of total energy is minimized using
the variational principle

\begin{equation}
\delta E[\Psi] = 0
\end{equation}

\begin{figure}[h]
\hspace{1pc}
\begin{minipage}{8pc}
\includegraphics[width=9pc]{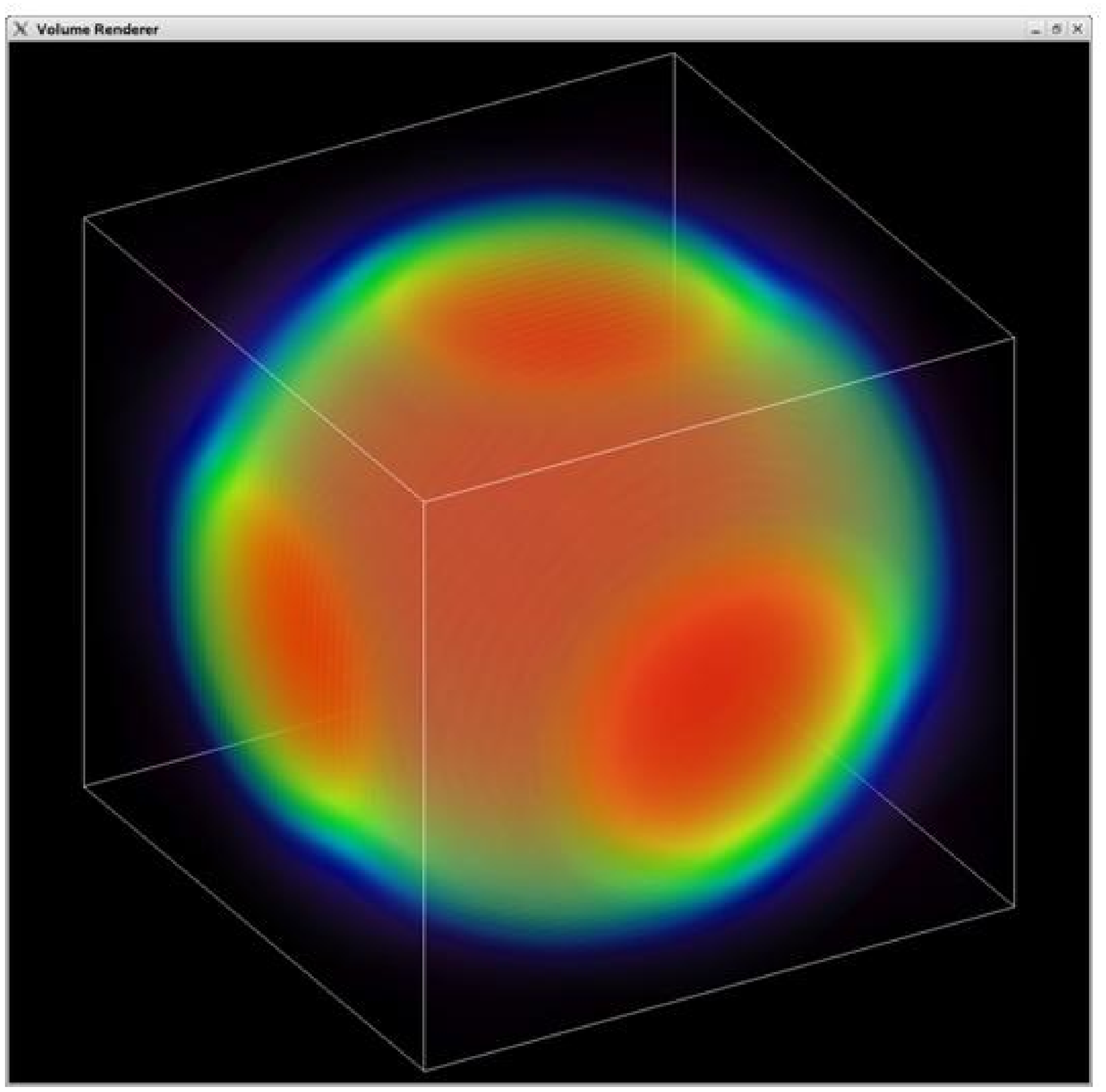}
\caption{\label{fig1}3D neutron density distribution after 500
iterations.}
\end{minipage}\hspace{1pc}%
\begin{minipage}{8pc}
\includegraphics[width=9pc]{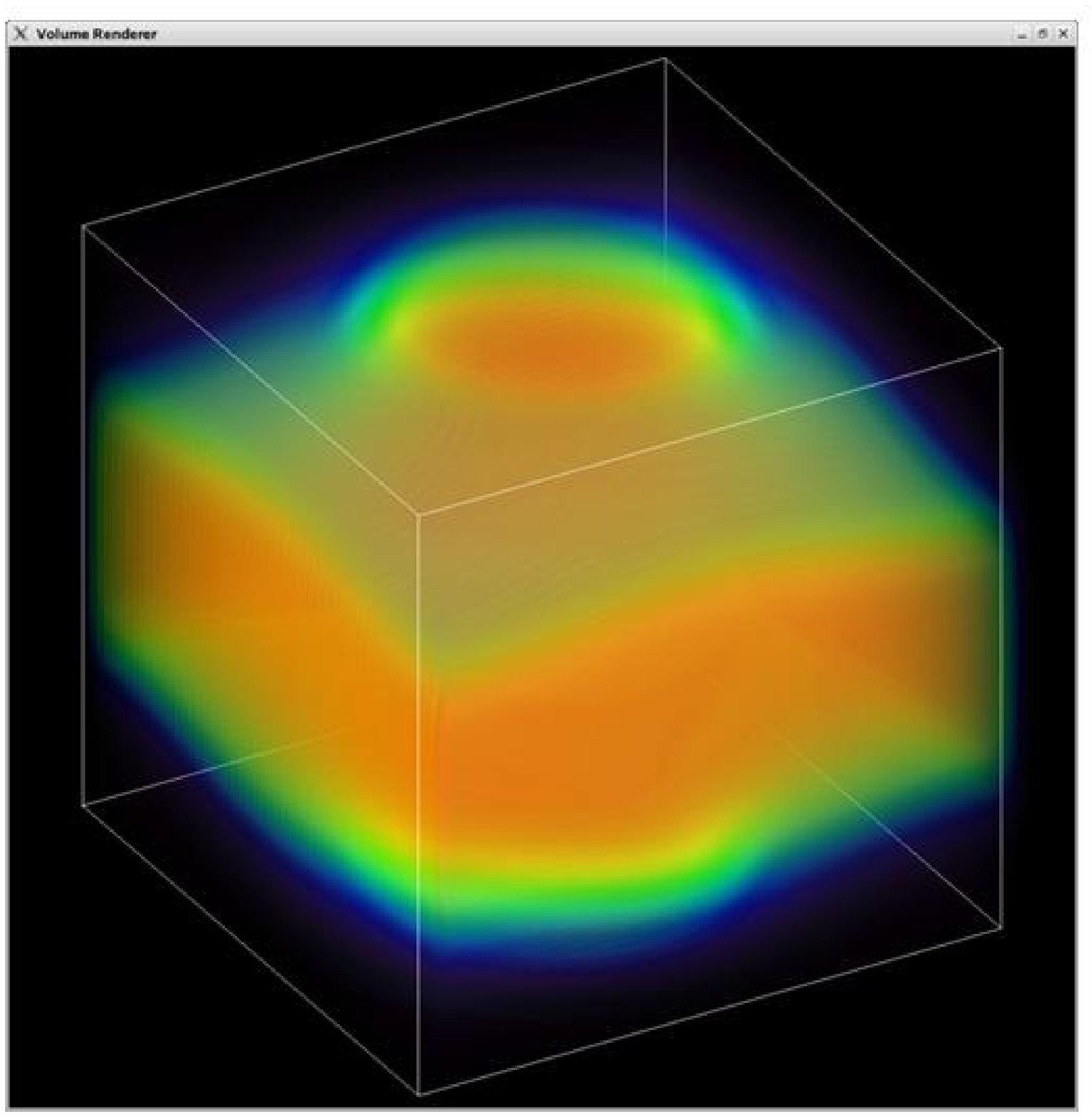}
\caption{\label{fig2}3D neutron density distribution after 2800
iterations.}
\end{minipage}\hspace{2pc}%
\begin{minipage}{8pc}
\includegraphics[width=9pc]{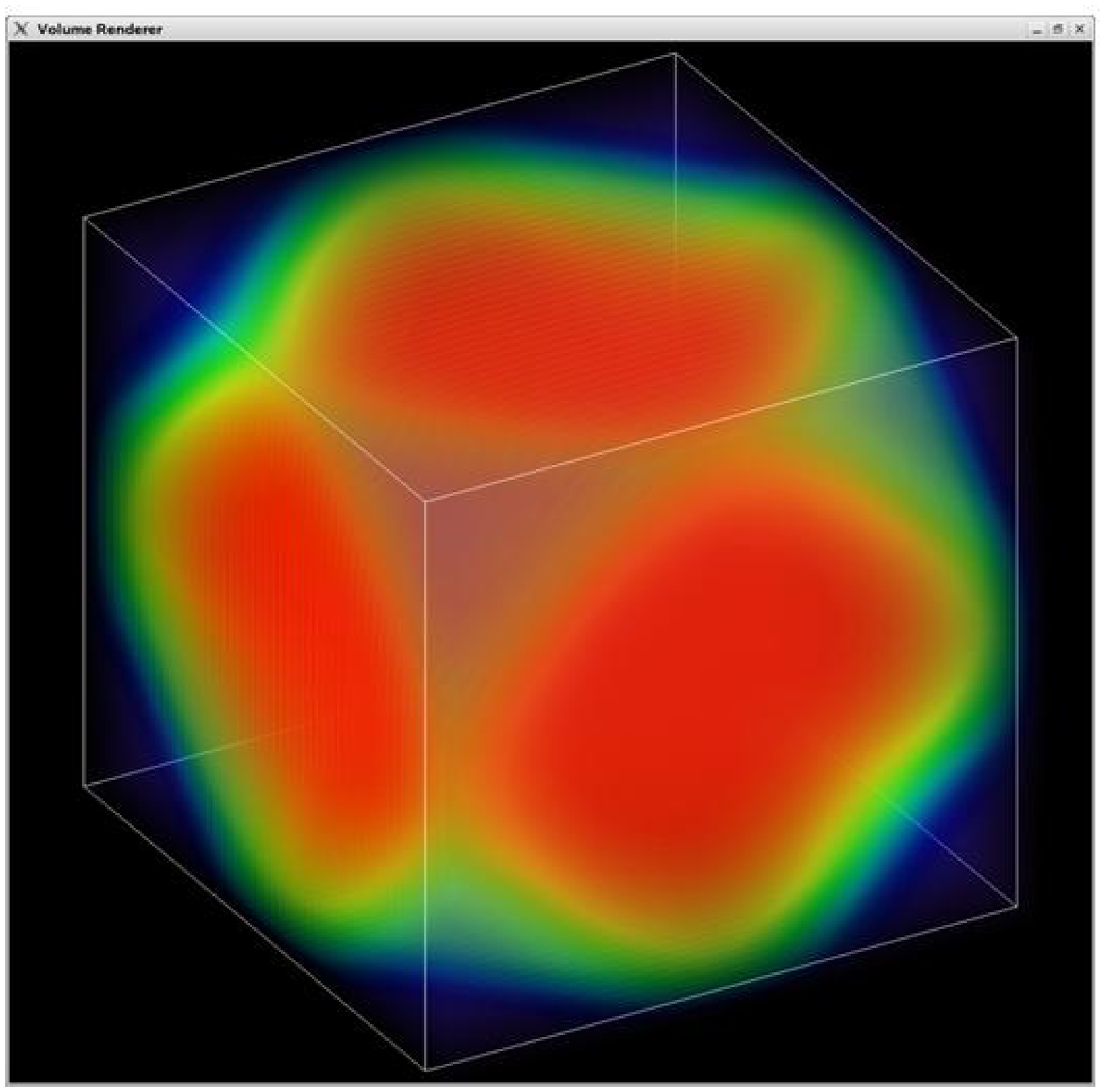}
\caption{\label{fig3}3D neutron density distribution after 500
iterations.}
\end{minipage}\hspace{1pc}%
\begin{minipage}{8pc}
\includegraphics[width=9pc]{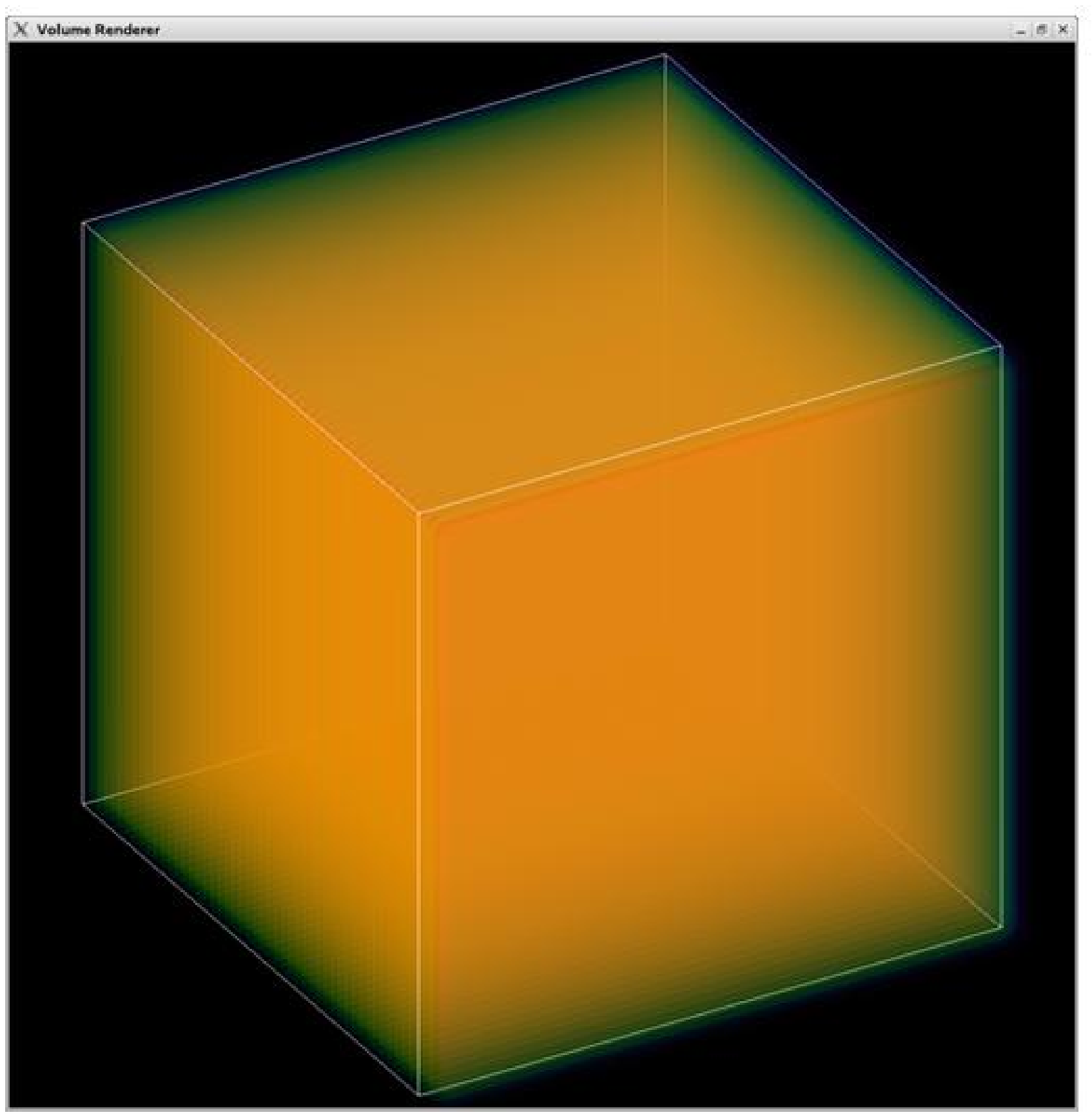}
\caption{\label{fig4}3D neutron density distribution after 6500
iterations.}
\end{minipage}
\end{figure}

This conditions leads to a system of A non-linear equations for
$\psi_i$ that has to be solved iteratively. In this work, three
forms of the trial wavefunction have been tested, Gaussian times
polynomial functions, harmonic oscillator wave functions and plane
waves. At the beginning the lowest A trial single-particle states
are occupied. After each iteration, the resulting states are
reordered according to increasing energy and re-occupied. This
approach ensures that the final solution is fully independent from
the initial choice of trial wavefunction and it is not
predetermined by this choice. The evolution of the shape of
neutron density distribution during the iteration process is
illustrated for A=900 and $y_{\rm p}$=0.3; in
Figs.~\ref{fig1}--\ref{fig2} the 3D density distribution is
displayed for $n_{\rm b}$=0.08 fm$^{\rm -3}$, $T$=2.5 MeV and
Figs.~\ref{fig3}--\ref{fig4} for $n_{\rm b}$=0.12 fm$^{\rm -3}$,
$T$=5.0 MeV. The change in the distribution after 500 and several
thousand iterations is quite striking. We note that in these
figures increase in density is color-coded from blue to red.

Two iteration schemes have been employed to avoid instabilities in
the iteration process - the Imaginary Time Step (ITS) and the
Damped Gradient Step (DGS). The ITS is very robust and leads to
initial rather rapid convergence even when the iteration process
is started from trial functions not too similar to the true
single-particle wavefunctions. However, when the minimum is
approached, it slows down exponentially. The DGS method requires
fairly good initial wavefunctions but converges much faster and
leads to close to linear convergence for final iterations. In the
present work both schemes have been used starting with the ITS and
switching over after first few hundred iterations to DGS. After
convergence is reached, the total energy density, entropy and
pressure and other related observables are calculated and the EoS
constructed in tabular form.

\begin{figure}[h]
\hspace{1pc}
\begin{minipage}{8pc}
\includegraphics[width=15pc]{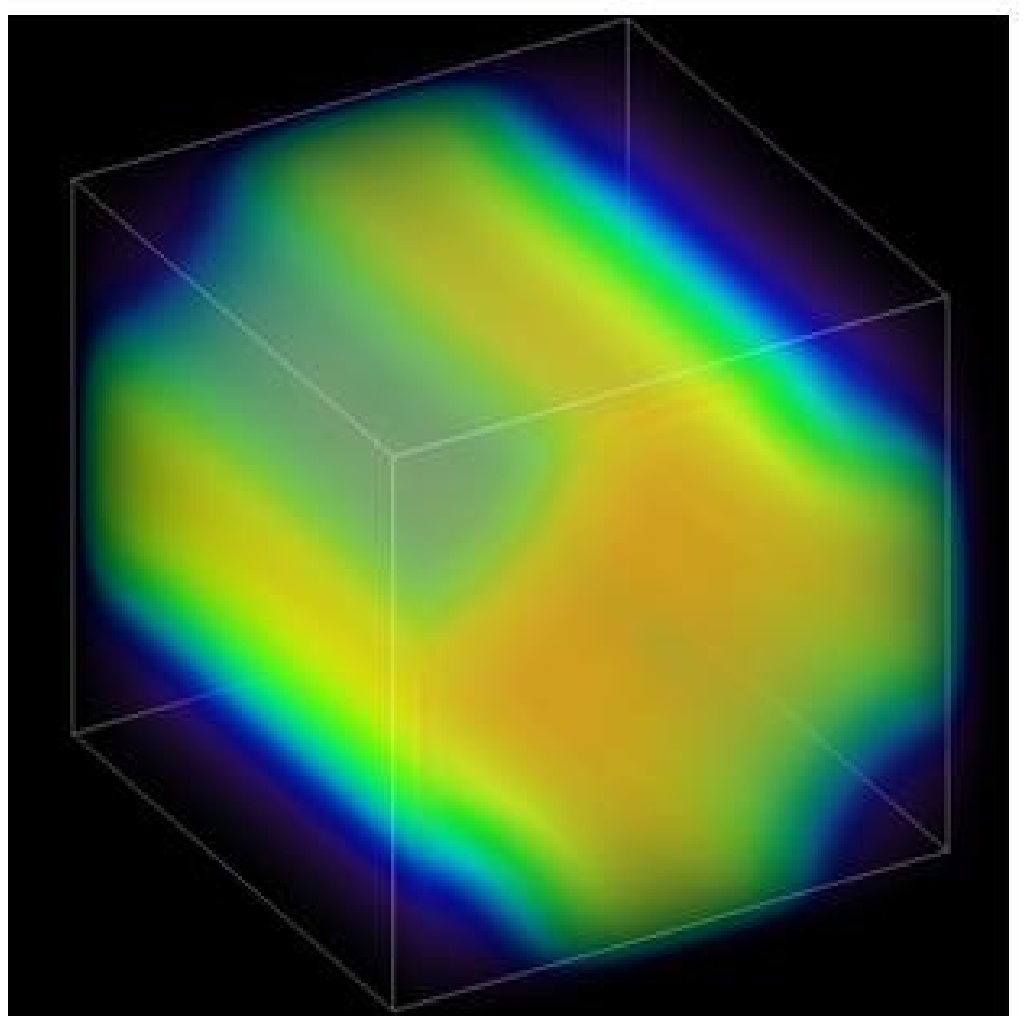}
\caption{\label{fig5}Neutron density distributions for A=180.}
\end{minipage}\hspace{1pc}%
\begin{minipage}{8pc}
\includegraphics[width=15pc]{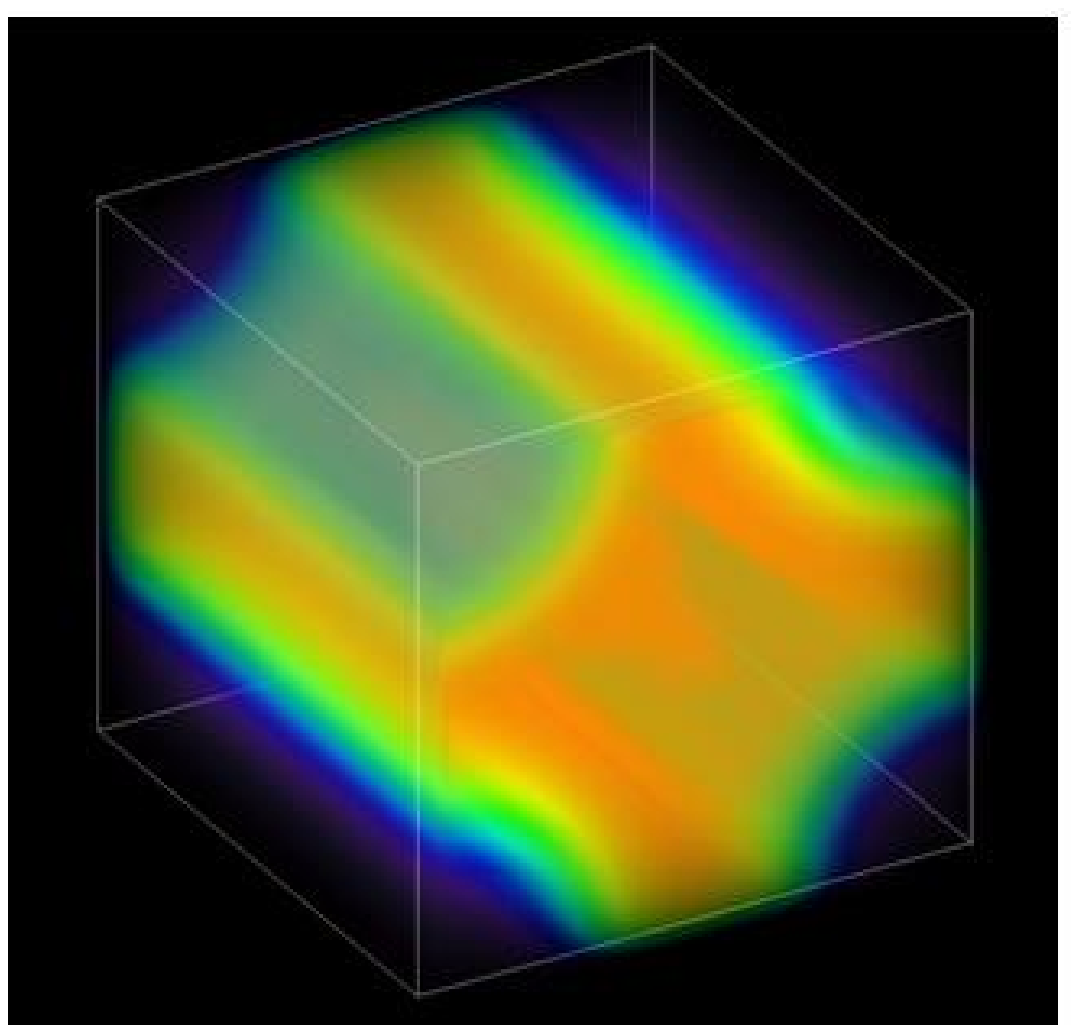}
\caption{\label{fig6}The same as Fig.~\protect\ref{fig5} but for
A=460.}
\end{minipage}\hspace{2pc}%
\begin{minipage}{8pc}
\includegraphics[width=15pc]{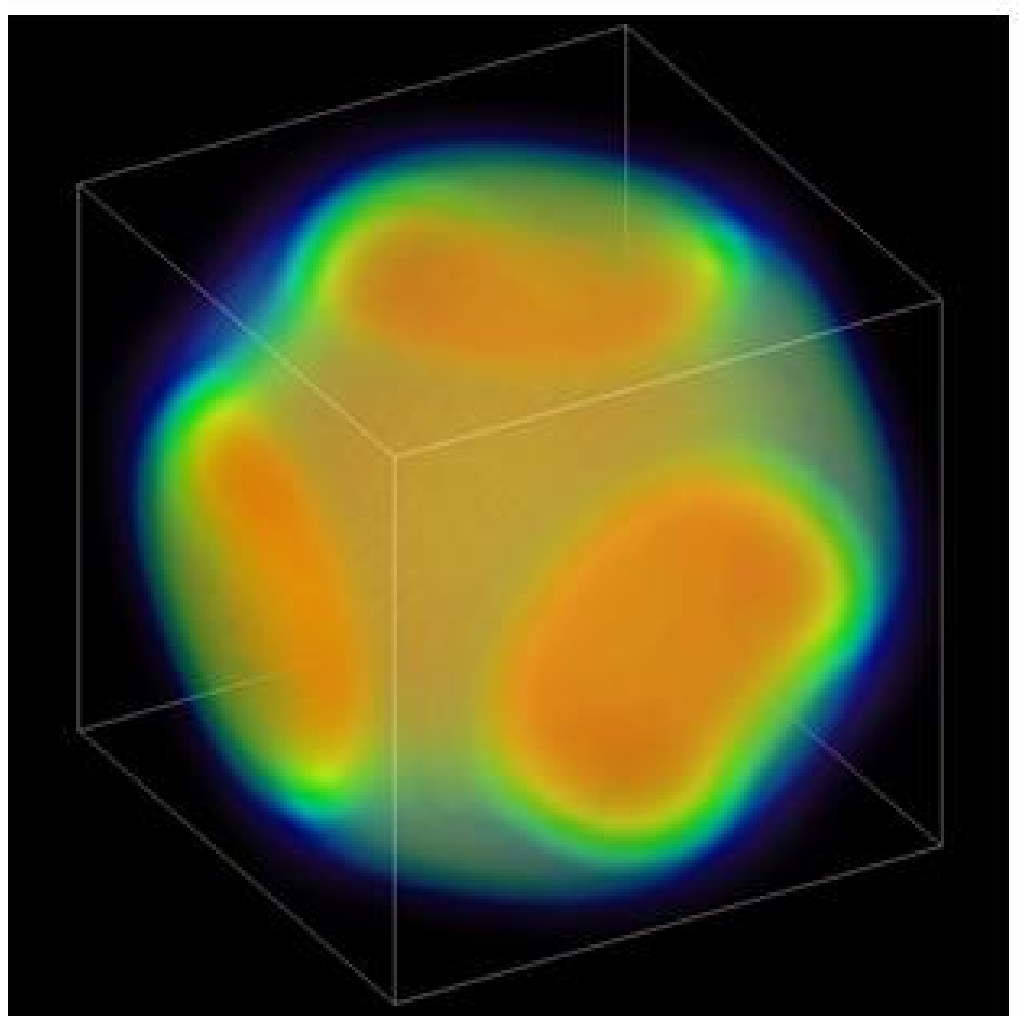}
\caption{\label{fig7}The same as Fig.~\protect\ref{fig5} but for
A=1400.}
\end{minipage}\hspace{1pc}%
\begin{minipage}{8pc}
\includegraphics[width=15pc]{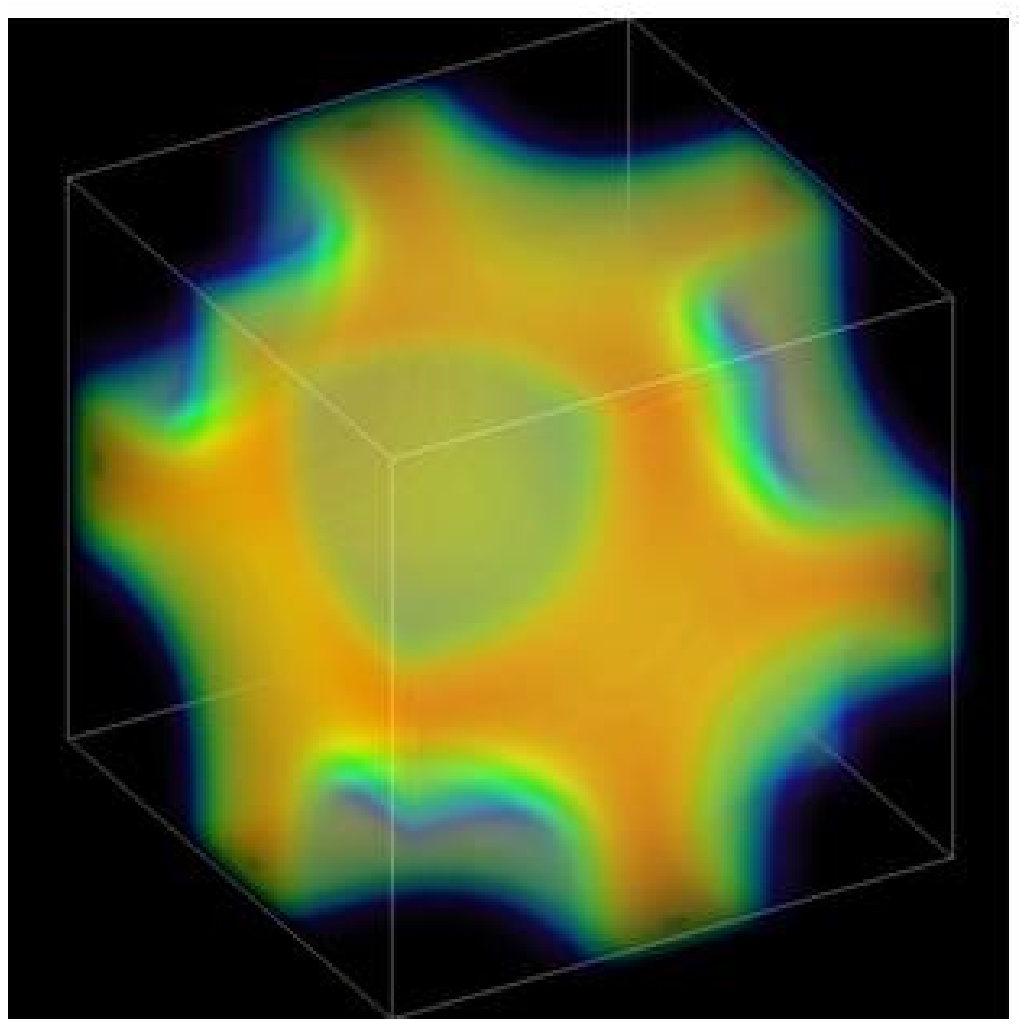}
\caption{\label{fig8}The same as Fig.~\protect\ref{fig5} but for
A=2200.}
\end{minipage}
\end{figure}

It is important to realise that it is not known \textit{ a priori}
what is the number of particles in the cell at given density that
corresponds to the physical size of the unit cell in nature. For
each particle number density the volume of a cell is defined as
$A/n_{\rm b}$ and the energy density and the spatial particle
density distribution varies significantly with $A$, as
demonstrated in Figs.~\ref{fig5}--\ref{fig8} for $n_{\rm b}$=0.08
fm$^{\rm -3}$, T=2.5 MeV and $y_{\rm p}$=0.3. Each of these
results are examples of possible \textit{excited} states of the
true unit cell (although they are local ground-states for a given
set of parameters). These states are rather close in energy and a
series of careful calculations has to be performed to search for
the value of $A$ which gives the absolute minimum energy density
for a given set of bulk parameters (i.e. \textit{minimum
minimorum}).

\section{Results and discussion}.

\begin{figure}[h]
\hspace{2pc}
\begin{minipage}{9pc}
\includegraphics[width=11pc]{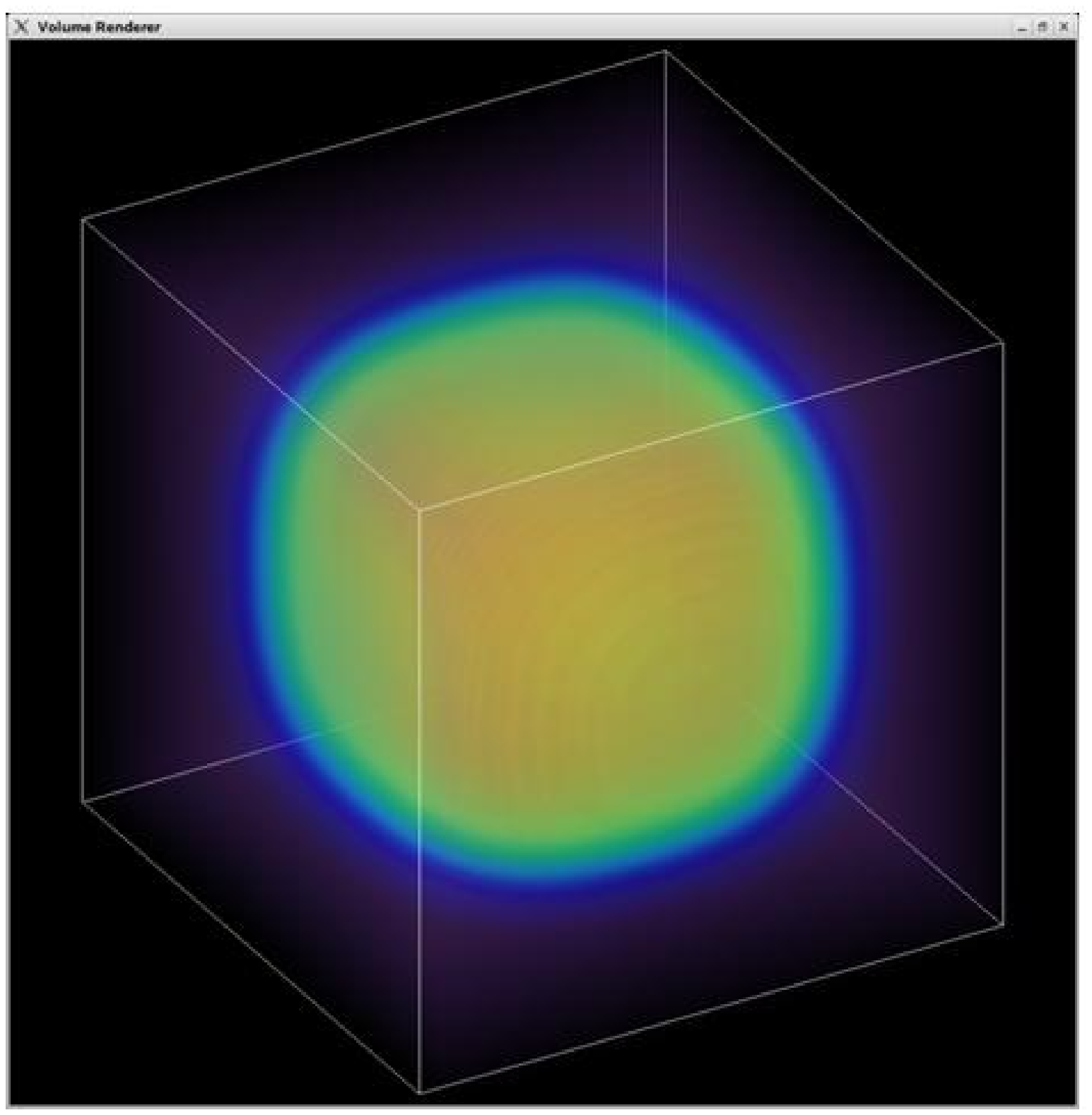}
\caption{\label{fig9}3D neutron density distribution at $n_{\rm
b}$=0.04 fm$^{\rm -3}$ .}
\end{minipage}\hspace{3pc}%
\begin{minipage}{9pc}
\includegraphics[width=11pc]{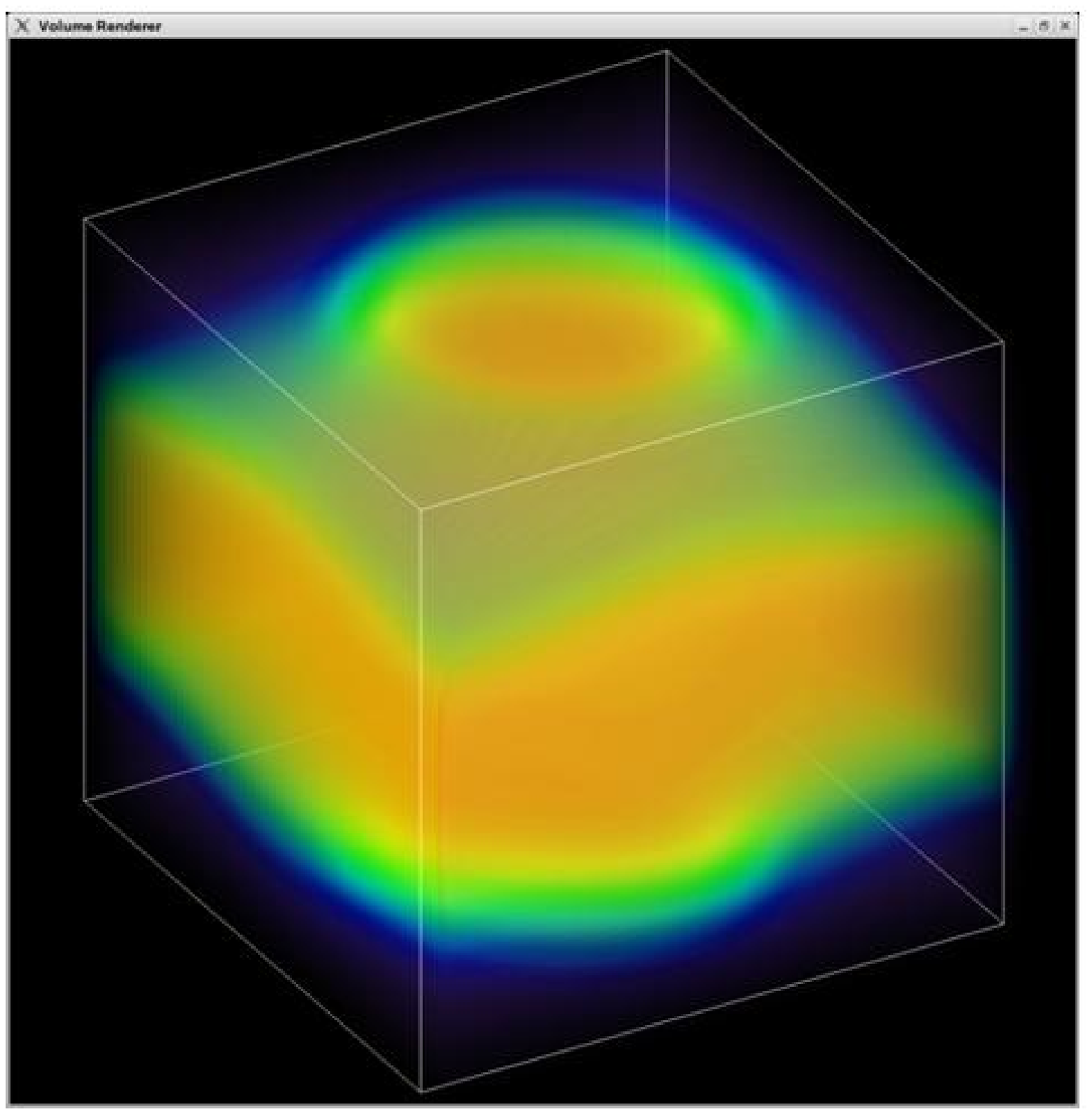}
\caption{\label{fig10} 3D neutron density distribution at $n_{\rm
b}$=0.08 fm$^{\rm -3}$.}
\end{minipage}\hspace{3pc}%
\begin{minipage}{9pc}
\includegraphics[width=11pc]{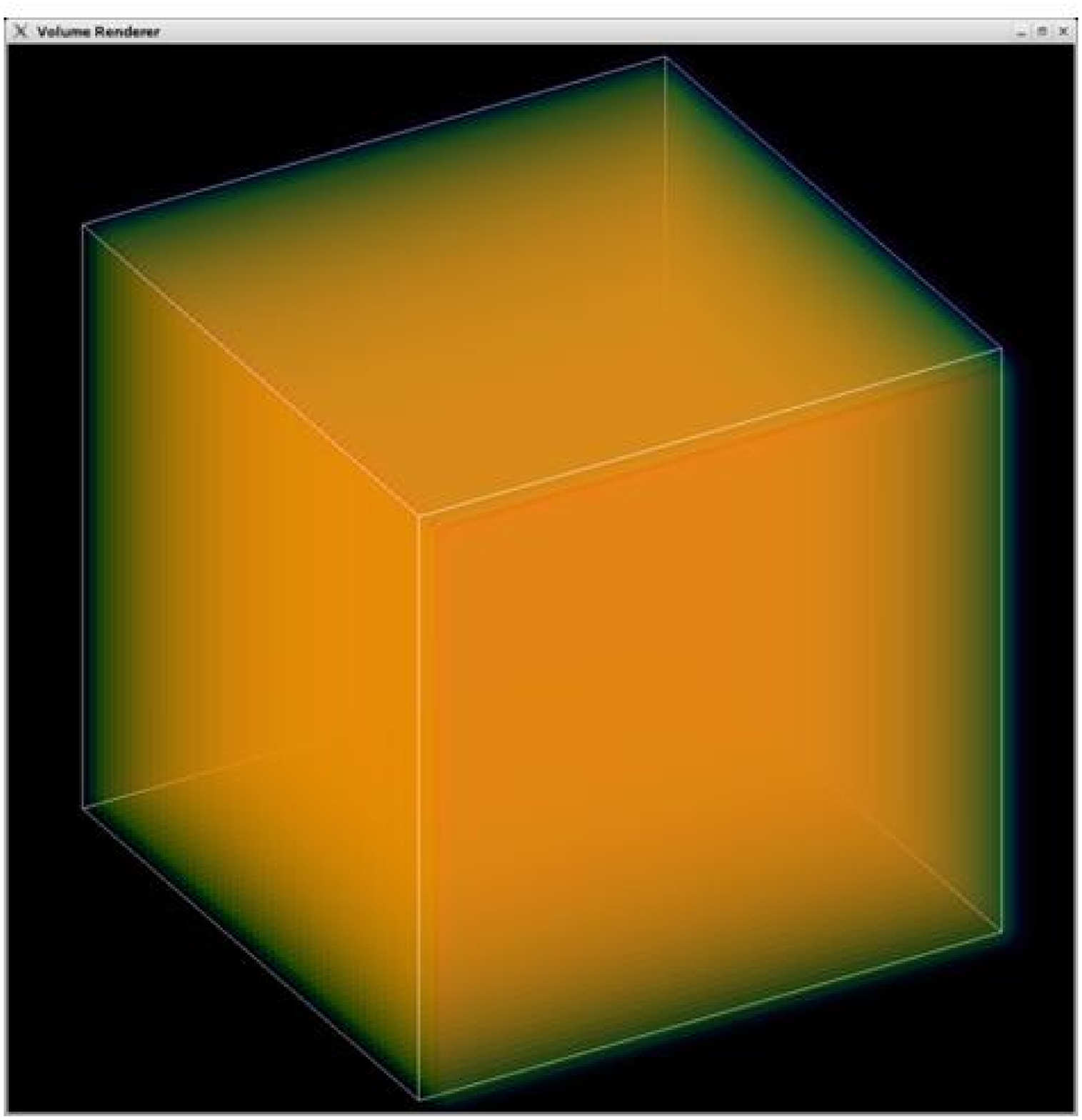}
\caption{\label{fig11} 3D neutron density distribution at $n_{\rm
b}$=0.12 fm$^{\rm -3}$.}
\end{minipage}
\end{figure}

One of the main results of the current work is the development of
properties of nuclear matter through extended density and
temperature regions. At the lower density limit ($n_{\rm b} <
0.0001$ fm$^{\rm -3}$), the nucleons are arranged as a roughly
spherical (but very large) `nucleus' at the centre of the cell. As
the density increases, however, the shape deforms and the nucleon
density distribution starts to spread out toward the cell
boundaries, assuming a variety of exotic forms made of high and
low density regions. At the extreme density, the nucleon density
distribution becomes uniform. This behaviour is illustrated in
Figs.~\ref{fig9}--\ref{fig11}  which shows neutron density
distribution at three selected densities, T=5 MeV and $y_{\rm
p}$=0.3 and clearly demonstrates the transition between spherical
and homogeneous density distribution.

The entire nucleon configuration within each cell is treated
self-consistently as one entity for each set of the macroscopic
parameters and evolves naturally within the model as the
macroscopic parameters are varied. This is in sharp contrast with
previous models where neutron heavy nuclei at and beyond the
particle drip-line where considered as immersed in a sea of
unbound free nucleons and the two systems were treated separately.
In this approach the transition between the inhomogeneous and
homogeneous phase of nuclear matter did not emerge naturally from
the calculation but had to be imposed artificially, introducing
uncertaintly about the threshold density region. Furthermore,
important phenomena discussed in more detail elsewhere
\cite{new06} such as shell effects, influence of the lattice
structure on Coulomb energy and scattering of weakly bound
nucleons on inhomogeneities in the matter are automatically
included.

\section{Summary}

The present model provides the first fully self-consistent 3D
picture of hot dense nuclear matter. It offers a new concept of
hot nuclear matter in the inner crust of neutron stars and in the
transitional density region between non-uniform and uniform matter
in collapsing stars. Instead of the traditional notion of
super-neutron-heavy nuclei immersed in a free neutron gas it
predicts a continuous medium with varying spatial concentration of
neutrons and protons. The properties of this medium come out
self-consistently from the model, as well as the transitions to
both higher and lower density phases of the matter. These results
may have profound consequences for macroscopic modelling of
core-collapse supernovae and neutron stars. In particular, weak
interaction processes (neutrino transport and beta-decay) in such
a  medium, will have to be investigated.

\ack  Special thanks go to R.~J.~Toede, Chao Li Wang, Amy Bonsor
and Jonathan Edge for development and performing data
visualisation. This work was conducted under the auspices of the
TeraScale Supernova Initiative, funded by SciDAC grants from the
DOE Office of Science High-Energy, Nuclear, and Advanced
Scientific Computing Research Programs and partly supported by US
DOE grant DE-FG02-94ER40834. Resources of the Center for
Computational Sciences at Oak Ridge National Laboratory were used.
Oak Ridge National Laboratory is managed by UT-Battelle, LLC, for
the U.S. Department of Energy under contract DE-AC05-00OR22725.

\hspace{4pc}

\newpage

\end{document}